\documentclass[letterpaper,prl,preprint,amsmath,amssymb,showpacs,superscriptaddress]{revtex4}
\usepackage[T1]{fontenc}
\usepackage{xcolor}
\usepackage{pdfcolmk}
\usepackage{prettyref}
\usepackage{amstext}
\usepackage{graphicx}
\usepackage{amssymb}
\PassOptionsToPackage{normalem}{ulem}
\usepackage{ulem}

\makeatletter

\providecolor{lyxadded}{rgb}{0,0,1}
\providecolor{lyxdeleted}{rgb}{1,0,0}

\@ifundefined{textcolor}{}
{%
 \definecolor{BLACK}{gray}{0}
 \definecolor{WHITE}{gray}{1}
 \definecolor{RED}{rgb}{1,0,0}
 \definecolor{GREEN}{rgb}{0,1,0}
 \definecolor{BLUE}{rgb}{0,0,1}
 \definecolor{CYAN}{cmyk}{1,0,0,0}
 \definecolor{MAGENTA}{cmyk}{0,1,0,0}
 \definecolor{YELLOW}{cmyk}{0,0,1,0}
 }

\makeatother

\begin{document}

\global\long\def\s{\:\mathrm{s}}
\global\long\def\ms{\:\mathrm{ms}}
\global\long\def\mV{\:\mathrm{mV}}
\global\long\def\Hz{\:\mathrm{Hz}}
\global\long\def\pA{\:\mathrm{pA}}
\global\long\def\pF{\:\mathrm{pF}}
\global\long\def\mus{\:\mu\mathrm{s}}
\global\long\def\Vth{V_{\theta}}
\global\long\def\Vr{V_{r}}
\global\long\def\PV{p(V)}
\global\long\def\Lfp{L_{\mathrm{FP}}}
\global\long\def\Qh{q_{h}}
\global\long\def\Qp{q_{p}}
\global\long\def\yr{y_{r}}
\global\long\def\yth{y_{\theta}}
\global\long\def\nuinst{\nu_{\mathrm{inst}}}
\global\long\def\cin{c_{\mathrm{in}}}
\global\long\def\cout{c_{\mathrm{out}}}
\global\long\def\Qy{q(y)}
\global\long\def\Q{q}
\global\long\def\erf{\mathrm{erf}}
\global\long\def\nue{\nu_{\mathrm{e}}}
\global\long\def\nui{\nu_{\mathrm{i}}}
\global\long\def\nuo{\nu_{0}}
\global\long\def\Presp{P_{\mathrm{r}}}
\global\long\def\erfc{\mathrm{erfc}}
\global\long\def\pinst{P_{\mathrm{inst}}}
\global\long\def\Jgamma{J(\gamma)}
\global\long\def\taur{\tau_{r}}
\global\long\def\defeq{\stackrel{\mathrm{def}}{=}}

\title{A Fokker-Planck formalism for diffusion with finite increments and
absorbing boundaries}

\author{Moritz Helias \footnote{Both authors contributed equally to this work.}}

\author{Moritz Deger \footnotemark[\value{footnote}]}

\author{Stefan Rotter}

\affiliation{Bernstein Center for Computational Neuroscience}

\affiliation{Computational Neuroscience, Faculty of Biology, Albert-Ludwig University\\
Hansastr. 9A, 79104 Freiburg, Germany}

\author{Markus Diesmann}

\affiliation{Bernstein Center for Computational Neuroscience}

\affiliation{RIKEN Brain Science Institute}

\affiliation{Brain and Neural Systems Team, RIKEN Computational Science Research
Program\\
2-1 Hirosawa, Wako City, Saitama 351-0198, Japan}

\date{November 13, 2009}

\begin{abstract}
Gaussian white noise is frequently used to model fluctuations in physical
systems. In Fokker-Planck theory, this leads to a vanishing probability
density near the absorbing boundary of threshold models. Here we derive
the boundary condition for the stationary density of a first-order
stochastic differential equation for additive finite-grained Poisson
noise and show that the response properties of threshold units are
qualitatively altered. Applied to the integrate-and-fire neuron model,
the response turns out to be instantaneous rather than exhibiting
low-pass characteristics, highly non-linear, and asymmetric for excitation
and inhibition. The novel mechanism is exhibited on the network level
and is a generic property of pulse-coupled systems of threshold units.

\end{abstract}
\maketitle

Dynamical systems driven by fluctuations are ubiquitous models in
solid state physics, quantum optics, chemical physics, circuit theory,
neural networks and laser physics. Absorbing boundaries are especially
interesting for diffusion over a potential step, escape from a particle
trap, or neuron models \cite{Stein65}. Approximating fluctuations
by Gaussian white noise enables analytical solutions by Fokker-Planck
theory \cite{Ricciardi79,Risken96}. For non-Gaussian noise, however,
the treatment of appropriate boundary conditions gains utmost importance
\cite{Hanggi85_1934}. The advantage of a transport description to
solve the mean first passage time problem has previously been demonstrated
\cite{VanDenBroeck84_2730}. In this line of thought, here we present
a novel hybrid theory that augments the classical diffusion approximation
by an approximate boundary condition for finite jump Poisson noise.
Exact results have so far only been obtained for the case of exponentially
distributed jumps amplitudes \cite{Jacobsen07_1330}. We apply our
theory to the leaky integrate-and-fire neuron model \cite{Stein65},
a noise-driven threshold system widely used to uncover the mechanisms
governing the dynamics of recurrent neuronal networks. An incoming
synaptic event causes a finite jump of the membrane potential which
decays back exponentially. The neuron fires a spike if the membrane
potential reaches a threshold. This simplicity renders the model analytically
tractable, efficient to simulate with precise spike times \cite{Morrison06c},
and yet it captures the gross features of neural dynamics \cite{Rauch03}.
The commonly pursued approach is to linearize this non-linear input-output
unit around a given level of background activity and to treat deviations
in the input as small perturbations. This technique has been applied
successfully to characterize the phase diagram of randomly connected
recurrent networks by a mean-field approach \cite{Brunel99}, to quantify
the transmission of correlated inputs by pairs of neurons \cite{DeLaRocha07_802,Tetzlaff02}
and to understand the interplay of spike-timing dependent learning
rules with neural dynamics \cite{Kempter99}. For Gaussian white noise
input current the linear response kernel is known exactly \cite{Brunel99}:
It constitutes a low-pass filter for signals modulating the mean \cite{Brunel01_2186};
only modulations of the fluctuations are transmitted immediately \cite{Lindner01_2934}.
In this Letter we show how our novel hybrid approach allows the analytical
prediction of a genuinely instantaneous non-linear response never
reported so far. Poisson noise with finite synaptic jumps even enhances
this response.

\section{Stochastic first order differential equation with finite increments}

We consider a first order stochastic differential equation driven
by point events from two Poisson processes with rates $\nu_{+}$ and
$\nu_{-}$. Each incoming event causes a finite jump $J_{k}=J_{+}>0$
for an increasing event and $J_{k}=J_{-}<0$ for a decreasing event\begin{eqnarray}
\dot{y} & = & f(y)+\sum_{k}J_{k}\delta(t-t_{k}).\label{eq:diffeq}\end{eqnarray}
We follow the notation in \cite{Ricciardi99} and employ the Kramers-Moyal
expansion with the infinitesimal moments $A_{n}(x,t)=\lim_{h\rightarrow0}\frac{1}{h}\langle(y(t+h)-y(t))^{n}\,|\, y(t)=x\rangle\quad n\in\mathbb{N}.$
The first and second infinitesimal moment evaluate to $A_{1}(x)=f(x)+\mu$
and $A_{2}=\sigma^{2},$ where we introduced the shorthand $\mu\defeq J_{+}\nu_{+}+J_{-}\nu_{-}$
and $\sigma^{2}\defeq J_{+}^{2}\nu_{+}+J_{-}^{2}\nu_{-}$. The time
evolution of the probability density $p(x,t)$ is then governed by
the Kramers-Moyal expansion, which we truncate after the second term
to obtain the Fokker-Planck equation \begin{align}
\frac{\partial}{\partial t}p(x,t) & =-\frac{\partial}{\partial x}\left[A_{1}(x)-\frac{1}{2}\frac{\partial}{\partial x}A_{2}\right]p(x,t)\label{eq:FP}\\
 & =-\frac{\partial}{\partial x}Sp(x,t),\nonumber \end{align}
where $S$ denotes the probability flux operator. In the presence
of an absorbing boundary at $\theta$, we need to determine the resulting
boundary condition for the stationary solution of \eqref{eq:FP}.
Without loss of generality, we assume an absorbing boundary at $\theta$
at the right end of the domain. Stationarity implies a constant flux
$\phi$ through the system. Rescaling the density by the flux as $q(x)=\phi^{-1}p(x)$
results in the linear inhomogeneous differential equation of first
order\begin{eqnarray}
Sq(x) & = & 1.\label{eq:diffeq_flux}\end{eqnarray}
The flux over the boundary has two contributions, the deterministic
drift and the positive stochastic jumps crossing the boundary \begin{eqnarray}
\phi & = & [f(\theta)]_{+}p(\theta)+\nu_{+}\pinst(J_{+})\label{eq:flux_explicit}\\
\pinst(s) & = & \int_{\theta-s}^{\theta}p(x)\, dx,\label{eq:P_inst}\end{eqnarray}
with $[x]_{+}=\{x\,\text{for }x>0;\,0\,\text{else}\}$. To evaluate
the integral in \eqref{eq:P_inst}, for small $J_{+}\ll\theta-\langle x\rangle$
we develop $q(x)$ into a Taylor series around $\theta$. To this
end, we solve \eqref{eq:diffeq_flux} for $q^{\prime}(x)=-\frac{2}{A_{2}}+\frac{2A_{1}(x)}{A_{2}}\, q(x)\defeq c_{1}+d_{1}(x)\, q(x).$
It is easy to see by induction, that all higher derivatives $q^{(n)}$
can be written in the form $q^{(n)}(x)=c_{n}(x)+d_{n}(x)q(x)$ whose
coefficients obey the recurrence relation \begin{eqnarray}
c_{n+1} & = & c_{n}^{\prime}+c_{1}d_{n}\qquad d_{n+1}=d_{n}^{\prime}+d_{1}d_{n}.\label{eq:recurrence_c_d}\end{eqnarray}
Inserting the Taylor series into \eqref{eq:P_inst} and performing
the integration results in\begin{eqnarray}
\pinst(s) & = & \sum_{n=0}^{\infty}\frac{1}{(n+1)!}\left.(c_{n}+d_{n}q)\right|_{\theta}(-s)^{n+1},\label{eq:P_inst_series}\end{eqnarray}
which is the probability mass moved across threshold by a perturbation
of size $s$ and hence also quantifies the instantaneous response
of the system. Dividing \eqref{eq:flux_explicit} by $\phi$ we solve
it for $q(\theta)$ to obtain the Dirichlet boundary condition\begin{eqnarray}
q(\theta) & = & \frac{1+\nu_{+}\sum_{n=0}^{\infty}\frac{1}{(n+1)!}c_{n}(\theta)(-J_{+})^{n+1}}{[f(\theta)]_{+}-\nu_{+}\sum_{n=0}^{\infty}\frac{1}{(n+1)!}d_{n}(\theta)(-J_{+})^{n+1}}.\label{eq:q_theta_explicit}\end{eqnarray}

If $J_{+}$ is small compared to the length scale on which the probability
density function varies, the probability density near the threshold
is well approximated by a Taylor polynomial of low degree; throughout
this letter, we truncate \eqref{eq:P_inst_series} and \eqref{eq:q_theta_explicit}
at $n=3$.

\section{Application to the leaky integrate-and-fire neuron}

We now apply the theory to a leaky integrate-and-fire neuron \cite{Stein65}
with membrane time constant $\tau$ and resistance $R$ receiving
excitatory and inhibitory synaptic inputs, as they occur in balanced
neural networks \cite{Vreeswijk96}. We model the input current $i(t)$
by point events $t_{k}\in\{\text{incoming spikes}\}$, drawn from
homogeneous Poisson processes with rates $\nue$ and $\nui$, respectively.
The membrane potential is governed by the differential equation $\tau\frac{dV}{dt}(t)=-V(t)+Ri(t)$.
An excitatory spike causes a jump of the membrane potential by $J_{k}=w$,
an inhibitory spike by $J_{k}=-gw$, so $Ri(t)=\tau\sum J_{k}\delta(t-t_{k})\,+Ri_{0}$,
where $i_{0}$ is a constant background current. Whenever $V$ reaches
the threshold $V_{\theta}$, the neuron emits a spike and the membrane
potential is reset to $V_{r}$, where it remains clamped for the absolute
refractory time $\tau_{r}$. With $\mu\defeq\tau w(\nu_{e}-g\nu_{i})$
and $\sigma^{2}\defeq\tau w^{2}(\nue+g^{2}\nui)$, we choose the natural
units $u=t/\tau$ and $y=(V-Ri_{0}-\mu)/\sigma$ to obtain $A_{1}(y)=-y$
and $A_{2}=1.$ The probability flux operator \eqref{eq:FP} is then
given as $S=-y-\frac{1}{2}\frac{\partial}{\partial y}.$ For the stationary
solution $q(y)=\nuo^{-1}p(y)$ the probability flux between reset
$\yr$ and threshold $\yth$ must equal the firing rate $\nuo$, and
is zero else, so \begin{eqnarray}
S\Q(y) & = & \begin{cases}
1 & \text{for }y_{r}\le y\le y_{\theta}\\
0 & \text{for }y<y_{r}.\end{cases}\label{eq:bc_flux}\end{eqnarray}

The equilibrium solution $\Q(y)=A\Qh(y)+\Qp(y)$ of \eqref{eq:bc_flux}
is a linear superposition of the homogeneous solution $\Qh(y)=e^{-y^{2}}$
and the particular solution $\Qp(y)=2e^{-y^{2}}\int_{\max(y_{r},y)}^{y_{\theta}}e^{u^{2}}\, du$,
chosen to be continuous at $y_{r}$ and to vanish at $\yth$. The
constant $A$ is determined from \eqref{eq:q_theta_explicit} as $A=\Q(\yth)/\Qh(\yth)$.
We obtain the mean firing rate $\nuo$ from the normalization condition
of the density $1=\nuo\tau\int_{-\infty}^{y_{\theta}}q(y)\, dy+\nu_{0}\taur$,
where $\nuo\taur$ is the fraction of neurons which are currently
refractory

\begin{align}
\frac{1}{\nuo}= & \tau\sqrt{\pi}\left[\int_{y_{r}}^{y_{\theta}}e^{y^{2}}(\erf(y)+1)\, dy+\frac{A}{2}(\mathrm{erf}(y_{\theta})+1)\right]+\taur.\label{eq:nu}\end{align}

\begin{figure}
\begin{centering}
\includegraphics[scale=0.8]{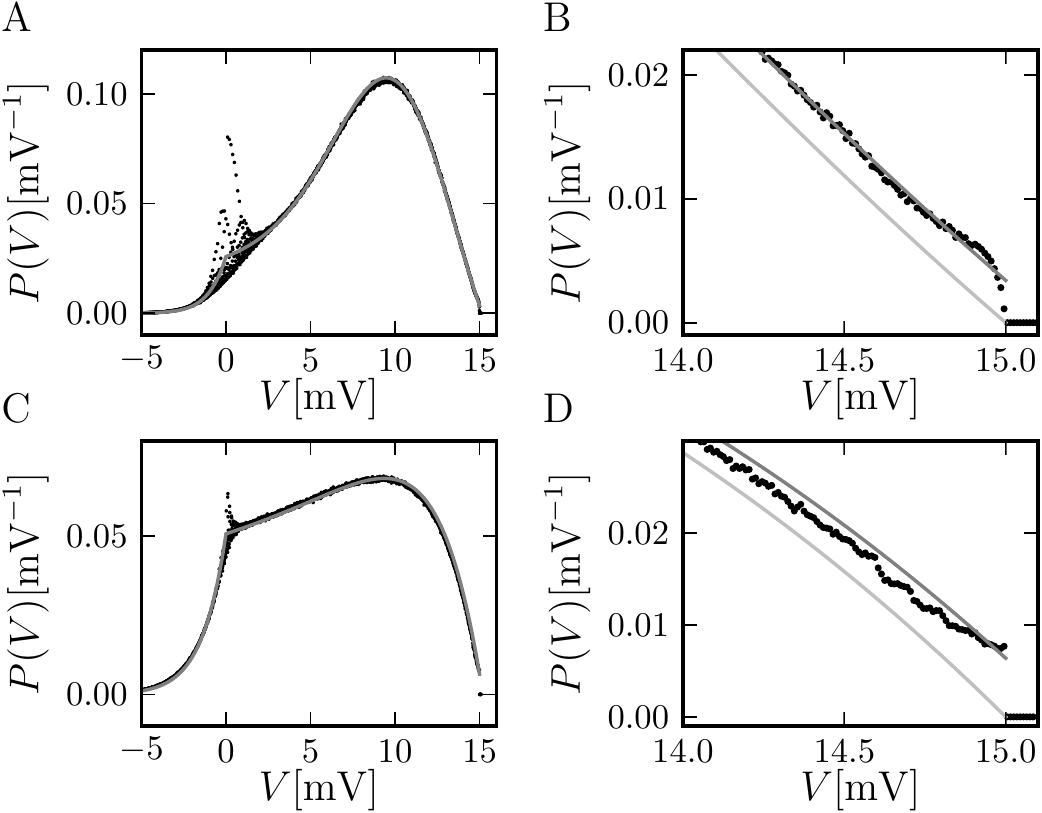}
\par\end{centering}

\caption{Finite synaptic potentials distort the stationary membrane potential
density $P(V)$. A Black: direct simulation. Parameters $\tau=20\ms$,
$V_{\theta}=15\mV$, $\Vr=0$, $i_{0}=0,$ $w=0.1\mV$, $g=4$, $\tau_{r}=1\ms$.
Incoming spike rates $\nu_{e}=29800\Hz$, $\nu_{i}=5950\Hz$ (corresponding
to $\mu=12\mV$ and $\sigma=5\mV$). Histogram binned with $\Delta V=0.01\mV$.
Gray: novel approximation $\nuo\tau/\sigma\Q((V-\mu-Ri_{0})/\sigma)$.
B Magnification of A around spike threshold. Light gray: solution
in diffusion limit of \cite{Brunel99}. C,D Density for supra-threshold
current $Ri_{0}=20\mV$ and incoming rates $\nue=95050\Hz$, $\nui=22262.5\Hz$
(corresponding to $\mu=12\mV$ and $\sigma=9.5\mV$). Other parameters
and gray code as in A,B.\label{fig:pdf_and_rate} }

\end{figure}

Figure \ref{fig:pdf_and_rate} shows the equilibrium solution near
the threshold obtained by direct simulation to agree much better with
our analytic approximation than with the theory for Gaussian white
noise input. Close to reset $V_{r}=0$, the oscillatory deviations
with periodicity $w$ are due to the higher occupation probability
for voltages that are integer multiples of a synaptic jump away from
reset. They wash out due to coupling of adjacent voltages by the deterministic
drift as one moves away from reset.

\begin{figure}
\begin{centering}
\includegraphics[scale=0.8]{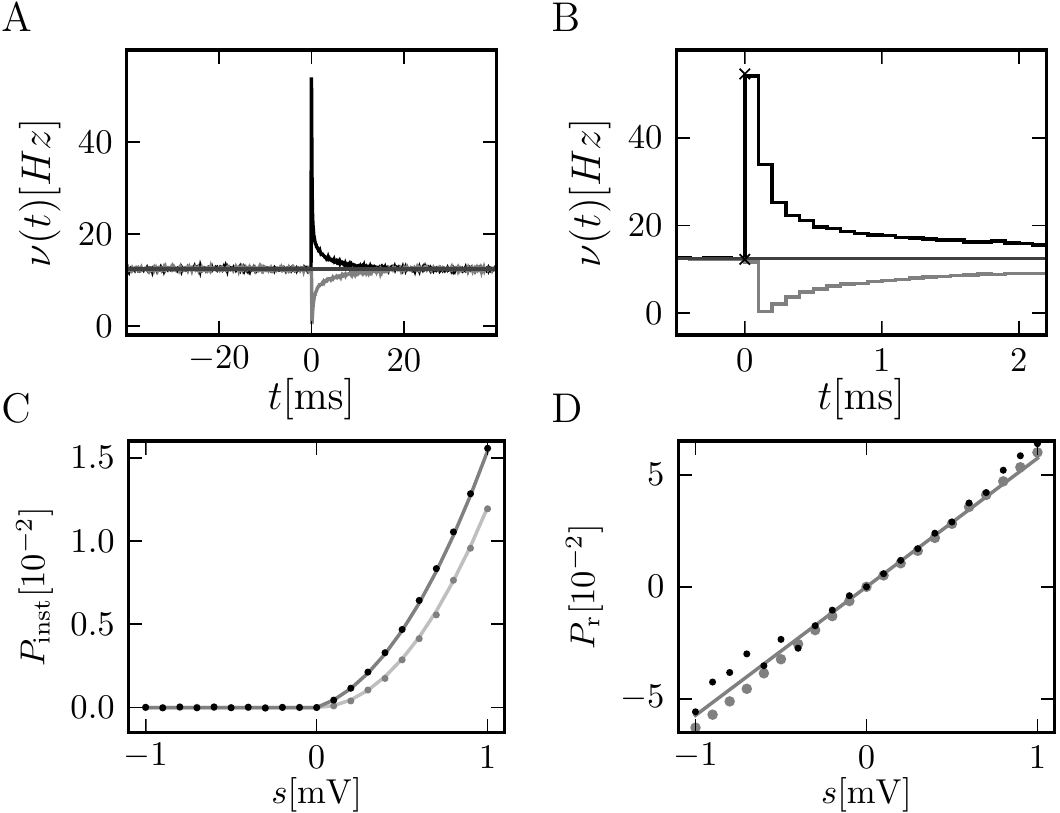}
\par\end{centering}

\caption{Instantaneous firing rate response to perturbation. A Black: Response
to perturbation $s=0.5\mV$ at $t=0$, gray: $s=-0.5\mV$. B Magnification
of A. Black cross: analytic peak response $\frac{\pinst}{h}$ \eqref{eq:P_inst_series}.
Histograms binned with $h=0.1\ms$. C Medium gray curve: instantaneous
response $\pinst$ \eqref{eq:P_inst_series} as a function of $s$
for finite weights $w=0.1\mV$. Black dots: Direct simulation. Light
gray curve: Diffusion limit of \eqref{eq:P_inst_series}. Medium gray
dots: Direct simulation of diffusion limit with temporal resolution
$10^{-4}\ms$. D Gray curve: integral response for finite weights
\eqref{eq:int_response}. Black dots: direct simulation. Gray dots:
direct simulation for Gaussian white noise background input. Simulated
data averaged over $2.5\cdot10^{8}\,(s=0.1\mV)\ldots2.5\cdot10^{6}\,(s=1.0\mV)$
perturbation events. Other parameters as in Figure \ref{fig:pdf_and_rate}A.}
\label{fig:response}
\end{figure}

We now proceed to obtain the response of the firing rate $\nu$ to
an additional $\delta$-shaped input current $\frac{\tau s}{R}\delta(t)$
fed into the neuron. This input causes a jump $s$ of the membrane
potential at $t=0$ and \eqref{eq:FP} suggests to treat it as a time
dependent perturbation of the mean input $\mu(t)=\mu+s\tau\delta(t)$.
First, we are interested in the integral response $\Presp(s)\defeq\int_{0}^{\infty}f_{s}(t)\, dt$
of the excess firing rate $f_{s}(t)=\nu_{s}(t)-\nuo$. Since the perturbation
has a flat spectrum, up to linear order in $s$ the spectrum of the
excess rate is $\hat{f_{s}}(z)=s\tau H(z)+O(s^{2})$, where $H(z)$
is the linear transfer function with respect to perturbing $\mu$
at Laplace frequency $z$. In particular, $P_{r}(s)=\hat{f_{s}}(0)$.
As $H(0)$ is the DC susceptibility of the system, we can express
it up to linear order as $H(0)=\frac{\partial\nuo}{\partial\mu}$.
Hence,

\begin{eqnarray}
\Presp(s) & = & \int_{0}^{\infty}\nu(t)-\nuo\, dt=s\tau\frac{d\nuo}{d\mu}+O(s^{2}).\label{eq:int_response}\end{eqnarray}
We also take into account the dependence of $A$ on $\mu$ to calculate
$\frac{d\nuo}{d\mu}$ from \eqref{eq:nu} and obtain

\begin{eqnarray}
\frac{d\nuo}{d\mu} & = & -\nuo^{2}\frac{\tau}{\sigma}\bigg(\sqrt{\pi}e^{\yr^{2}}\erfc(-\yr)-Q(\yth)\nonumber \\
 &  & +\erfc(-\yth)\left(\frac{\Q(\yth)-\Q(\yth-\frac{w}{\sigma})}{\erf(\yth)-\erf(\yth-\frac{w}{\sigma})}\right)\bigg)\;.\label{eq:d_nu_d_mu}\end{eqnarray}
Figure \ref{fig:response}D shows the integral response to be in good
agreement with the linear approximation. The integral response in
the diffusion limit is almost identical.

The instantaneous response of the firing rate to an impulse-like perturbation
can be quantified without further approximation. The perturbation
shifts the probability density by $s$ so that neurons with $V\in[\Vth-s,\Vth${]}
immediately fire. This results in the finite firing probability $\pinst(s)$
of the single neuron within infinitesimal time \eqref{eq:P_inst},
which is zero for $s<0$. This instantaneous response has several
interesting properties: For small $s$ it can be approximated in terms
of the value and the slope of the membrane potential distribution
below the threshold (using \eqref{eq:P_inst_series} for $n\le2$),
so it has a linear and a quadratic contribution in $s$. Figure \ref{fig:response}A
shows a typical response of the firing rate to a perturbation. The
peak value for a positive perturbation agrees well with the analytic
approximation \eqref{eq:P_inst} (Figure \ref{fig:response}C).

\begin{figure}
\begin{centering}
\includegraphics[scale=0.8]{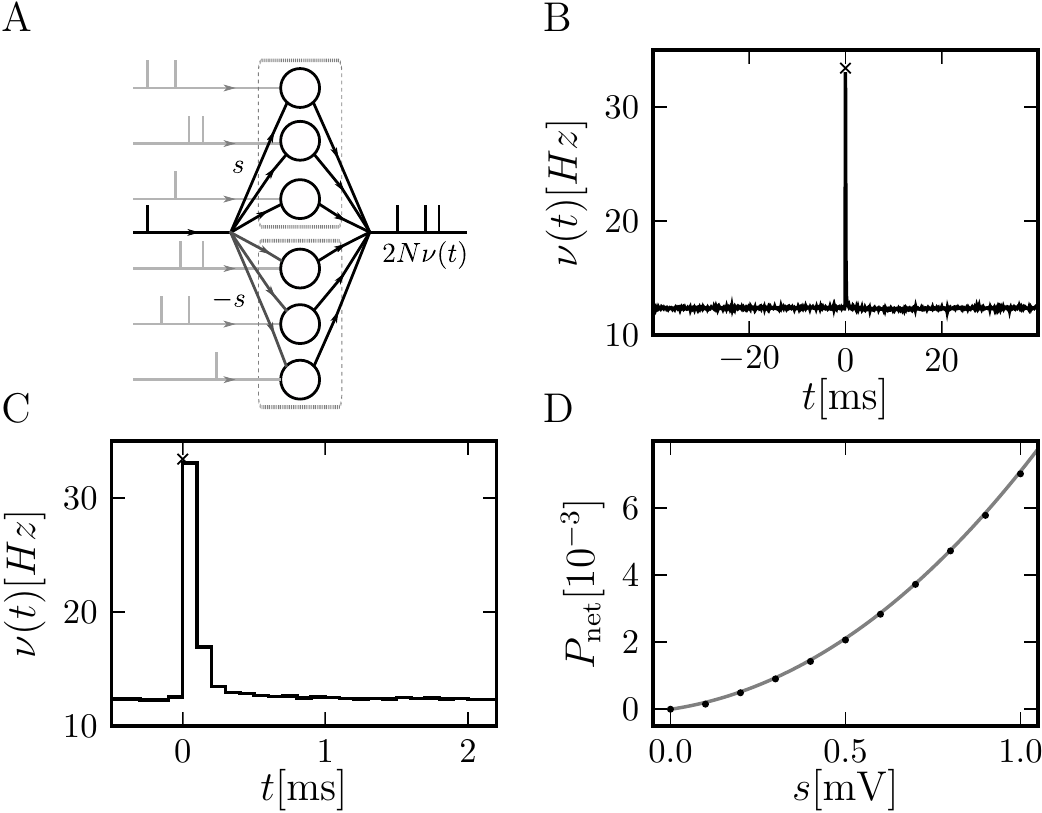}
\par\end{centering}

\caption{The non-linear response to perturbations is exhibited on the network
level. A Two identical populations of $N=1000$ neurons each, receive
uncorrelated background input (light gray spikes). At $t=0$ the neurons
simultaneously receive an additional input of size $s$ in the upper
and $-s$ in the lower population (symbolized by black single spike).
B Pooled response of the populations normalized by the number of neurons.
C Magnification of B. D Instantaneous response $P_{\mathrm{net}}(s)$
(black dots: direct simulation, gray curve: analytical result) as
a function of $s$. Other parameters as in Figure \ref{fig:response}A.}
\label{fig:linear_response_canceling}
\end{figure}

Due to the expansive nature of the instantaneous response (Figure
\ref{fig:response}C) its relative contribution to the integral response
increases with $s$. For realistic synaptic weights $\le1\mV$ the
contribution reaches $\simeq30$ percent. Replacing the background
input by Gaussian white noise, and using the boundary condition $q(\yth)=0$
in \eqref{eq:P_inst_series} yields a smaller instantaneous response
(Figure \ref{fig:response}C) which for positive $s$ still exhibits
a quadratic, but no linear, dependence. The integral response, however,
does not change significantly (Figure \ref{fig:response}D). An example
network in which the linear non-instantaneous response cancels completely
and the instantaneous response becomes dominant is shown in Figure
\ref{fig:linear_response_canceling}A. At $t=0$ two populations of
neurons simultaneously receive a perturbation of size $s$ and $-s$
respectively. This activity may, for example, originate from a third
pool of synchronous excitatory and inhibitory neurons. The pooled
linear firing rate response of the former two populations is zero.
The instantaneous response, however, causes a very brief overshoot
at $t=0$ (Figure \ref{fig:linear_response_canceling}B). Figure \ref{fig:linear_response_canceling}C
reveals that the response returns to baseline within$\simeq0.3\ms$.
Figure \ref{fig:linear_response_canceling}D shows the non-linear
dependence of peak height $P_{\mathrm{net}}(s)=\frac{1}{2}\pinst(s)$
on $s$.

In this Letter we present an extension to the diffusion approximation
with finite but small increments. Our theory describes the probability
density evolution as a diffusion on the length scale $\sigma$ determined
by the fluctuations, but we take the quantization of the noise near
an absorbing boundary into account, leading to a non-zero Dirichlet
condition. This hybrid approach enables us to find analytical approximations
hitherto unknown for pure jump models. In particular we accurately
quantify the instantaneous contribution to the escape rate in response
to a perturbation. There is a formal similarity to the findings of
\cite{Goedeke08_015007} for large perturbations. Applied to the integrate-and-fire
neuron with Gaussian white noise input, we quantify a quadratic non-linear
instantaneous firing rate response not captured by the existing linear
theory \cite{Brunel01_2186,Lindner01_2934}. Finite jumps in the noise
qualitatively change and enhance the response compared to the case
of Gaussian white noise: for small perturbations, the quadratic dependence
is now dominated by an additional linear term.  The instantaneous
and the integral response both display stochastic resonance (not shown,
see \cite{helias09_epaps}) as observed for periodic perturbations
\cite{Lindner01_2934} and for aperiodic stimuli in adiabatic approximation
\cite{Collins96c}. The results in this Letter are obtained by integrating
the neural dynamics in continuous time; simulations in discrete time
exaggerate the instantaneous response \cite{Helias09a_pre}. The
diffusion approximation still limits our approach: for weights $w\ge0.2\mV$
higher moments than order two neglected by the Fokker-Planck equation
become relevant \cite{helias09_epaps}. Also, the oscillatory modulations
of the probability density on the scale $w\ll\sigma$ in the regions
below threshold and around the reset potential are outside the scope
of our theory. In a different approach restricted to excitatory finite
sized inputs, \cite{sirovich00_2009} calculated the equilibrium solution.
It remains to be checked whether our results extend to biologically
more realistic spike-onset mechanisms \cite{Fourcaud03_11640,Naundorf05_297}.
The novel effect is exhibited on the macroscopic level and we expect
it to contribute to synchronization phenomena and the correlation
transmission in physical systems.
\begin{acknowledgments}
We acknowledge fruitful discussions with Carl van Vreeswijk, Nicolas
Brunel, and Benjamin Lindner and are grateful to our colleagues in
the NEST Initiative. Partially funded by BMBF Grant 01GQ0420 to BCCN
Freiburg, EU Grant 15879 (FACETS), DIP F1.2, Helmholtz Alliance on
Systems Biology (Germany), and Next-Generation Supercomputer Project
of MEXT (Japan).

\end{acknowledgments}

\end{document}


\global\long\def\s{\:\mathrm{s}}
\global\long\def\ms{\:\mathrm{ms}}
\global\long\def\mV{\:\mathrm{mV}}
\global\long\def\Hz{\:\mathrm{Hz}}
\global\long\def\pA{\:\mathrm{pA}}
\global\long\def\pF{\:\mathrm{pF}}
\global\long\def\mus{\:\mu\mathrm{s}}
\global\long\def\Vth{V_{\theta}}
\global\long\def\Vr{V_{r}}
\global\long\def\PV{P(V)}
\global\long\def\Lfp{L_{\mathrm{FP}}}
\global\long\def\Qh{q_{h}}
\global\long\def\Qp{q_{p}}
\global\long\def\yr{y_{r}}
\global\long\def\yth{y_{\theta}}
\global\long\def\nuinst{\nu_{\mathrm{inst}}}
\global\long\def\cin{c_{\mathrm{in}}}
\global\long\def\cout{c_{\mathrm{out}}}
\global\long\def\Qy{q(y)}
\global\long\def\Q{q}
\global\long\def\erf{\mathrm{erf}}
\global\long\def\nue{\nu_{\mathrm{e}}}
\global\long\def\nui{\nu_{\mathrm{i}}}
\global\long\def\nuo{\nu_{0}}
\global\long\def\Presp{P_{\mathrm{r}}}
\global\long\def\erfc{\mathrm{erfc}}
\global\long\def\pinst{P_{\mathrm{inst}}}
\global\long\def\Jgamma{J(\gamma)}
\global\long\def\taur{\tau_{r}}
\global\long\def\defeq{\stackrel{\mathrm{def}}{=}}

\title{A Fokker-Planck formalism for diffusion with finite increments\\
and absorbing boundaries\\
Supplementary material}

\author{Moritz Helias \footnote{Both authors contributed equally to this work.}}

\author{Moritz Deger \footnotemark[\value{footnote}]}

\author{Stefan Rotter}

\affiliation{Bernstein Center for Computational Neuroscience}

\affiliation{Computational Neuroscience, Faculty of Biology, Albert-Ludwig University\\
Hansastr. 9A, 79104 Freiburg, Germany}

\author{Markus Diesmann}

\affiliation{Bernstein Center for Computational Neuroscience}

\affiliation{RIKEN Brain Science Institute}

\affiliation{Brain and Neural Systems Team, RIKEN Computational Science Research
Program\\
2-1 Hirosawa, Wako City, Saitama 351-0198, Japan}

\date{November 13, 2009}

\maketitle

\section{Stochastic first order differential equation with finite jumps}

We consider a dynamic variable governed by a differential equation
of first order with Poissonian input causing finite jumps $J_{k}=J_{+}>0$
for increasing events arriving with rate $\nu_{+}$ and $J_{k}=J_{-}<0$
for decreasing events of rate $\nu_{-}$\begin{eqnarray}
\dot{y} & = & f(y)+\sum_{k}J_{k}\delta(t-t_{k}).\label{eq:diffeq}\end{eqnarray}
We follow the notation in \cite{Ricciardi99} and employ the Kramers
Moyal expansion with the infinitesimal moments\begin{eqnarray*}
A_{n}(x,t) & = & \lim_{h\rightarrow0}\frac{1}{h}\langle(y(t+h)-y(t))^{n}\,|\, y(t)=x\rangle\quad n\in\mathbb{N}.\end{eqnarray*}
 The first infinitesimal moment evaluates to

\begin{align}
A_{1}(x)= & \lim_{h\rightarrow0}\frac{1}{h}\mathrm{\langle}y(t+h)-y(t)\,|\, y(t)=x\rangle\nonumber \\
= & \lim_{h\rightarrow0}\frac{1}{h}\langle x+hf(x)+\underbrace{J_{+}\kappa_{+}+J_{-}\kappa_{-}}_{\iota(t,h)}-x\rangle\nonumber \\
= & f(x)+J_{+}\nu_{+}+J_{-}\nu_{-}\nonumber \\
= & f(x)+\mu\label{eq:A1}\end{align}
where $\kappa_{+}$, and $\kappa_{-}$ denote the Poisson distributed
number of positive and negative incoming events in the time interval
of size $h$, respectively. We used $\lim_{h\rightarrow0}\frac{1}{h}\kappa_{+}=\nu_{+}$,
$\lim_{h\rightarrow0}\frac{1}{h}\kappa_{-}=\nu_{-}$ and introduced
the shorthand $\mu\defeq J_{+}\nu_{+}+J_{-}\nu_{-}$. The second infinitesimal
moment is\begin{align*}
A_{2}(x,t)= & \lim_{h\rightarrow0}\frac{1}{h}\langle(y(t+h)-y(t))^{2}\,|\, y(t)=x\rangle\\
= & \lim_{h\rightarrow0}\frac{1}{h}\left\langle \left(x+hf(x)+\iota(t,h)-x\right)^{2}\right\rangle \\
= & \lim_{h\rightarrow0}\frac{1}{h}\left(h^{2}f^{2}(x)+2hf(x)\,\langle\iota(t,h)\rangle+\mathrm{\langle}\iota(t,h){}^{2}\rangle\right)\\
= & \lim_{h\rightarrow0}\frac{1}{h}\mathrm{\langle}\iota(t,h){}^{2}\rangle,\end{align*}
where the first and second term in the second last row vanish for
$h\rightarrow0$ and for the independent individual Poisson distributed
random numbers $\kappa_{+}$ and $\kappa_{-}$ it holds\begin{eqnarray*}
\langle(J_{+}\kappa_{+}+J_{-}\kappa_{-})^{2}\rangle & = & J_{+}^{2}(\nu_{+}h+(\nu_{+}h)^{2})+J_{-}^{2}(\nu_{-}h+(\nu_{-}h)^{2})+2J_{+}J_{-}h^{2}\nu_{+}\nu_{-}\\
 & = & h(J_{+}^{2}\nu_{+}+J_{-}^{2}\nu_{-})+h^{2}(J_{+}\nu_{+}+J_{-}\nu_{-})^{2}\end{eqnarray*}
So the infinitesimal variance is independent of $x$\begin{eqnarray}
A_{2}= & J_{+}^{2}\nu_{+}+J_{-}^{2}\nu_{-} & \defeq\sigma^{2}\label{eq:A2}\end{eqnarray}
The time evolution of the density is then governed by the Kramers-Moyal
expansion \begin{align}
\frac{\partial}{\partial t}p(x,t)= & \sum_{n=1}^{\infty}\frac{1}{n!}\left(-\frac{\partial}{\partial x}\right)^{n}A_{n}(x,t)\, p(x,t).\label{eq:fokker_planck_series}\end{align}
We truncate this series after the second term to obtain the Fokker-Planck
equation \begin{align}
\frac{\partial}{\partial t}p(x,t) & =\Lfp p(x,t)\nonumber \\
 & =-\frac{\partial}{\partial x}\left[A_{1}(x)-\frac{1}{2}\frac{\partial}{\partial x}A_{2}(x)\right]p(x,t)\\
 & =-\frac{\partial}{\partial x}Sp(x,t),\label{eq:FP}\end{align}
with the probability flux operator $S=A_{1}(x)-\frac{1}{2}\frac{\partial}{\partial x}A_{2}(x)=f(x)+\mu-\frac{\sigma^{2}}{2}\frac{\partial}{\partial x}$.

\section{Absorbing Boundary condition\label{sec:bc_and_solution}}

Without loss of generality, we assume an absorbing bondary at $\theta$
that has to be crossed from below. We need to determine the proper
boundary condition for the stationary solution $\Lfp p(x)=0$ at the
threshold. Stationarity implies a constant flux $\phi$ through the
system. Rescaling the solution by the flux as $q(x)=\phi^{-1}p(x)$
results in the linear inhomogeneous differential equation of first
order\begin{eqnarray}
Sq(x) & = & 1.\label{eq:diffeq_flux}\end{eqnarray}
On the other hand, the flux crossing the boundary from left has two
contributions, the deterministic drift and the positive stochastic
jumps crossing the boundary \begin{eqnarray}
\phi & = & [f(\theta)]_{+}p(\theta)+\nu_{+}\pinst(J_{+})\label{eq:flux_explicit}\\
\pinst(s) & = & \int_{\theta-s}^{\theta}p(x)\, dx,\label{eq:P_inst}\end{eqnarray}
with $[x]_{+}=\{x\,\text{for }x>0;\,0\,\text{else}\}$. For small
$J_{+}\ll\theta-\langle x\rangle$ we express $q(x)$ near the threshold
$\theta$ by a Taylor series. To this end, we solve \eqref{eq:diffeq_flux}
for $q^{\prime}(x)=-\frac{2}{A_{2}}+\frac{2A_{1}(x)}{A_{2}}\, q(x)\defeq c_{1}+d_{1}(x)\, q(x).$
It is easy to see by induction, that all higher derivatives $q^{(n)}$
can be written in the form $q^{(n)}(x)=c_{n}(x)+d_{n}(x)q(x)$ which
yields the recurrence relations \begin{eqnarray}
c_{n+1} & = & c_{n}^{\prime}+c_{1}d_{n}\label{eq:recurrence_c_d}\\
d_{n+1} & = & d_{n}^{\prime}+d_{1}d_{n}.\nonumber \end{eqnarray}
Inserting the Taylor series of $q$ around $\theta$ \begin{eqnarray}
q(x) & = & \sum_{n=0}^{\infty}\frac{1}{n!}\left.(c_{n}+d_{n}q)\right|_{\theta}(x-\theta)^{n}\label{eq:Q_approx}\end{eqnarray}
into \eqref{eq:P_inst} and integrating results in\begin{eqnarray}
\pinst(s) & = & \sum_{n=0}^{\infty}\frac{1}{(n+1)!}\left.(c_{n}+d_{n}q)\right|_{\theta}(-s)^{n+1}.\label{eq:P_inst_series}\end{eqnarray}
Inserted into \eqref{eq:flux_explicit} and dividing by $\phi$ we
solve for $q(\theta)$ to obtain the Dirichlet boundary condition\begin{eqnarray}
q(\theta) & = & \frac{1+\nu_{+}\sum_{n=0}^{\infty}\frac{1}{(n+1)!}c_{n}(\theta)(-J_{+})^{n+1}}{[f(\theta)]_{+}-\nu_{+}\sum_{n=0}^{\infty}\frac{1}{(n+1)!}d_{n}(\theta)(-J_{+})^{n+1}}.\label{eq:q_theta_explicit}\end{eqnarray}

We can make use of the fact, that typically $J_{+}$ is small compared
to the length scale on which the probability density function varies.
So the probability density near the threshold is well approximated
by a Taylor polynomial of low degree; throughout this letter, we truncate
\eqref{eq:P_inst_series} and \eqref{eq:q_theta_explicit} at $n=3$.

\section{Leaky integrate-and-fire neuron}

We now apply the theory to the leaky integrate-and-fire neuron \cite{Stein65}
with membrane time constant $\tau$ and resistance $R$ receiving
excitatory and inhibitory synaptic inputs, as they occur in balanced
neural networks \cite{Vreeswijk96}. We model the input current $i(t)$
by point events $t_{k}\in\{\text{incoming spikes}\}$, drawn from
homogeneous Poisson processes with rates $\nue$ and $\nui$, respectively.
The membrane potential is governed by the differential equation

\begin{eqnarray*}
\tau\frac{dV}{dt}(t) & = & -V(t)+Ri(t)\\
Ri(t) & = & \tau\sum J_{k}\delta(t-t_{k})\,+Ri_{0};\end{eqnarray*}
an excitatory spike causes a jump of the membrane potential by $J_{k}=w$,
an inhibitory by $J_{k}=-gw$. Whenever $V$ reaches the threshold
$V_{\theta}$, the neuron emits a spike and the membrane potential
is reset to $V_{r}$ where it remains clamped for the absolute refractory
time $\tau_{r}$. $i_{0}$ is a constant input current. With $\mu\defeq w\tau(\nue-g\nui)$
and $\sigma\defeq w^{2}\tau(\nue+g^{2}\nui)$ we choose the natural
units $u=t/\tau$ and $y=\frac{V-Ri_{0}-\mu}{\sigma}$ proposed by
\eqref{eq:A1} and \eqref{eq:A2} to obtain the form \eqref{eq:diffeq}\begin{eqnarray*}
\frac{\partial y}{\partial u} & = & \underbrace{-y-\frac{\mu}{\sigma}}_{f(y)}+\sum_{k}\frac{J_{k}}{\sigma}\delta(u-u_{k}),\end{eqnarray*}
 where the events $u_{k}=t_{k}/\tau$ arrive with rate $\tau\nue$
(excitatory) and $\tau\nui$, respectively. Consequently, in these
coordinates $A_{1}(y)=-y$ and $A_{2}=1$, so the probability flux
operator \eqref{eq:FP} acting on the rescaled probability density
$q(y)=\nuo^{-1}p(y)$ is given as \begin{align}
S= & -y-\frac{1}{2}\frac{\partial}{\partial y}.\label{eq:flux_dimless_cont}\end{align}
 For the stationary solution $q$ of \eqref{eq:FP}, the probability
flux between reset and threshold must equal the stationary firing
rate \begin{eqnarray}
S\Qy & = & \begin{cases}
1 & \text{for }y_{r}\le y\le y_{\theta}\\
0 & \text{for }y<y_{r}.\end{cases}\label{eq:bc_flux}\end{eqnarray}
 The equilibrium solution of \eqref{eq:bc_flux} is a linear superposition
of the homogeneous solution $\Qh$ satisfying the differential equation
$S\Qh=0$ given by \eqref{eq:flux_dimless_cont}, which is

\begin{align}
\Qh(y)= & e^{-y^{2}},\label{eq:Q_h}\end{align}
and the particular solution $\Qp$. The latter is chosen to be continuous
at $y_{r}$, to vanish at $\yth$, and to fulfill the inhomogeneous
differential equation \eqref{eq:bc_flux} $\Qp^{\prime}=-2y\Qp-2H(y-y_{r})$
(with $H$ denoting the Heaviside function)

\begin{align}
\Qp(y)= & -Q_{h}(y)\int_{\yth}^{y}Q_{h}^{-1}(u)2H(u-\yr)\, du\label{eq:Q_p}\\
= & 2e^{-y^{2}}\int_{\max(y_{r},y)}^{y_{\theta}}e^{u^{2}}\, du\;.\end{align}
The solution which fulfills the boundary conditions is the superposition\begin{align}
\Q= & \Qp+A\,\Qh\;,\label{eq:Q_full}\end{align}
where the constant $A$ is determined as $A=q(y_{\theta})\Qh^{-1}(\yth)$.
Using the recurrence \eqref{eq:recurrence_c_d} we determine the sequence
of the $c_{n}$ and $d_{n}$ as \begin{eqnarray*}
\{c_{n}(y)\} & = & \{-2,4y,-8y^{2}+8,\ldots\}\\
\{d_{n}(y)\} & = & \{-2y,4y^{2}-2,-8y^{3}+12y,\ldots\}.\end{eqnarray*}

The Dirichlet value at the threshold of given by \eqref{eq:q_theta_explicit}
as

\begin{eqnarray*}
q(\theta) & = & \frac{1+\tau\nu_{e}\sum_{n=0}^{\infty}\frac{1}{(n+1)!}c_{n}(\theta)(-\frac{w}{\sigma})^{n+1}}{[-\frac{\Vth-Ri_{0}}{\sigma}]_{+}-\tau\nu_{e}\sum_{n=0}^{\infty}\frac{1}{(n+1)!}d_{n}(\theta)(-\frac{w}{\sigma})^{n+1}}.\end{eqnarray*}
Since the explicit solution \eqref{eq:Q_h} of $\Qh$ is known and
since we have chosen $\Qp(\yth)=0$, it follows that $d_{n}(\yth)e^{-\yth^{2}}$
is the $n$-th derivative $\Qh^{(n)}(\yth)$ so that we can replace
the sum in the denominator by\begin{eqnarray*}
-e^{-\yth^{2}}\sum_{n=0}^{\infty}\frac{1}{(n+1)!}d_{n}(\theta)(-\frac{w}{\sigma})^{n+1} & = & \int_{\yth-\frac{w}{\sigma}}^{\yth}e^{-y^{2}}\, dy\\
 & = & \frac{\sqrt{\pi}}{2}(\erf(\yth)-\erf(\yth-\frac{w}{\sigma})).\end{eqnarray*}
Hence the boundary value can be calculated as\begin{eqnarray*}
A & = & \frac{1+\tau\nu_{e}\sum_{n=0}^{\infty}\frac{1}{(n+1)!}c_{n}(\theta)(-\frac{w}{\sigma})^{n+1}}{e^{-\yth^{2}}[-\frac{\Vth-Ri_{0}}{\sigma}]_{+}+\tau\nu_{e}\frac{\sqrt{\pi}}{2}(\erf(\yth)-\erf(\yth-\frac{w}{\sigma}))}.\end{eqnarray*}
The mean firing rate $\nuo$ can be obtained from the normalization
of $\Q$, which reads $1=\int_{-\infty}^{\theta}\PV dV+\nu_{0}\taur=\tau\nuo\int_{-\infty}^{y_{\theta}}\Qy dy+\nuo\taur$.
The term $\nuo\taur$ is the fraction of neurons which are currently
refractory. We obtain \begin{align}
(\tau\nuo)^{-1} & =\int_{-\infty}^{\yth}A\Qh(y)+\Qp(y)\, dy+\tau_{r}/\tau\nonumber \\
\int_{-\infty}^{\yth}\Qh(y)\, dy & =\frac{\sqrt{\pi}}{2}(\erf(\yth)+1)\nonumber \\
\int_{-\infty}^{\yth}\Qp(y)\, dy & =2\int_{-\infty}^{\yr}e^{-y^{2}}\int_{y_{r}}^{y_{\theta}}e^{u^{2}}\, du+2\int_{\yr}^{\yth}e^{-y^{2}}\int_{y}^{y_{\theta}}e^{u^{2}}\, du\nonumber \\
 & =\sqrt{\pi}(1+\erf(\yth))\int_{y_{r}}^{y_{\theta}}e^{u^{2}}\, du+\left.\sqrt{\pi}(1+\erf(y))\int_{y}^{y_{\theta}}e^{u^{2}}\, du\right|_{y=\yr}^{y=\yth}+\sqrt{\pi}\int_{\yr}^{\yth}(1+\erf(y))e^{y^{2}}\, dy\nonumber \\
 & =\sqrt{\pi}\int_{\yr}^{\yth}(1+\erf(y))e^{y^{2}}\, dy\nonumber \\
\nuo^{-1}= & \tau\sqrt{\pi}\left(\frac{1}{2}A(\mathrm{erf}(y_{\theta})+1)+\int_{y_{r}}^{y_{\theta}}e^{y^{2}}(\erf(y)+1)\, dy\right)+\taur.\label{eq:nu}\end{align}

\begin{figure}
\begin{centering}
\includegraphics{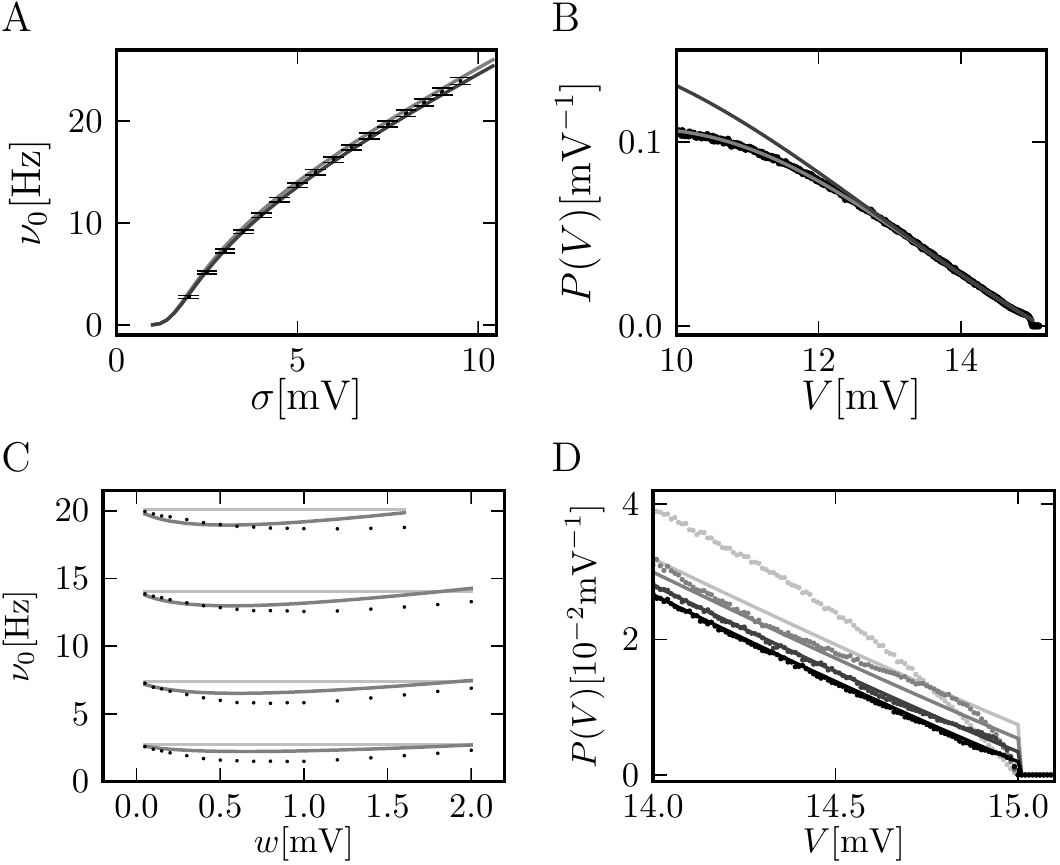}
\par\end{centering}

\caption{A analytic firing rate compared to simulation depending on fluctuations
$\sigma$ of the input. Black: Analytic solution \eqref{eq:nu}, light
gray: firing rate in diffusion limit \cite{Siegert51,Brunel00_183}.
Parameters $\tau=20\ms$, $V_{\theta}=15\mV$, $\Vr=0$, $w=0.1\mV$,
$g=4$, $\tau_{r}=1\ms$. $\nu_{e}$, $\nu_{i}$ chosen to realize
mean input $\mu=12\mV$ and the fluctuation $\sigma$ as given by
the abscissa. B Probability density near threshold. Black dots: direct
simulation, solid gray line: analytic result $P(V)=\nuo\tau/\sigma Q((V-\mu)/\sigma)$
given by \eqref{eq:Q_full}. Dark gray solid line: Polynomial of degree
$3$ approximating the density around $\Vth$ using \eqref{eq:Q_approx}.
Parameters as in A, but with $\nu_{e}=29800\Hz$, $\nu_{i}=5950\Hz$
(corresponding to $\mu=12\mV$ and $\sigma=5\mV$). C Analytic firing
rate compared to simulation depending on the size of synaptic jumps
$w$ for fixed $\sigma=5\mV$ and $\mu=[7,9.5,12,14]\mV$ (from bottom
to top). Gray code as in A. D Membrane potential distribution near
threshold for $\mu=12\mV$ and $\sigma=5\mV$ for different $w=0.05\mV$
(black) $w=0.1\mV$ (dark gray), $w=0.2\mV$ (gray), $w=0.5\mV$ (light
gray). }
\label{fig:poly_approx_Q}
\end{figure}

A comparison of this approximation to the full solution is shown in
Figure \ref{fig:poly_approx_Q}C. Within the range of several synaptic
weights $w=0.1\mV$ used here, the approximation is nearly perfect.

\section{Response properties\label{sec:response_properties}}

In this section, we calculate the response of an integrate-and-fire
neuron to an additional $\delta$-shaped input current. We follow
two approaches here: First, we quantify the integral response in linear
perturbation theory. Then we go beyond linear response and determine
the non-linear instantaneous response to a $\delta$-shaped input
current.

Treating the neuron in linear response theory, it is sufficient to
determine the response of the neuron's firing rate with respect to
an input $\delta$-current at $t=0$. Such a perturbation causes a
jump of the membrane potential and can be treated as a time dependent
perturbation of the mean input by $\mu(t)=\mu+s\tau\delta(t)$ as
can be seen from \eqref{eq:diffeq}. We call the time-dependent response
of the firing rate$\nu_{s}(t)$ and its excess over baseline $f_{s}(t)=\nu_{s}(t)-\nuo$.
The integral response is defined as $\Presp(s):=\int_{0}^{\infty}f_{s}(t)\, dt$.
We denote by $\hat{f_{s}}(z)=\int_{0}^{\infty}e^{zt}f_{s}(t)\, dt$
its Laplace transform. The spectrum of the perturbation is flat, so
to linear order in $s$ the spectrum of the response is $\hat{f_{s}}(z)=s\tau H(z)+O(s^{2})$,
where $H(z)$ is the linear transfer function of the system with respect
to perturbing $\mu$. In particular, $P_{r}(s)=\hat{f_{s}}(0)$. Since
$H(0)$ is the linear DC susceptibility of the system (changing $\mu$
infinitesimally slowly), we can equate $H(0)=\frac{\partial\nuo}{\partial\mu}$.
Hence we arrive at

\begin{eqnarray}
\Presp(s) & = & \int_{0}^{\infty}\nu(t)-\nuo\, dt=s\tau\frac{d\nuo}{d\mu}+O(s^{2}).\label{eq:int_response}\end{eqnarray}
We calculate $\frac{d\nuo}{d\mu}$, from \eqref{eq:nu}, where we
need to take into account also the dependence of $A$ on $\mu$. With
$\frac{d\nu_{0}^{-1}}{d\mu}=-\frac{1}{\nu_{0}^{2}}\frac{d\nu_{0}}{d\mu}$,
we only need to calculate $\frac{d\nu_{0}^{-1}}{d\mu}$. Additionally,
we note that $\frac{dy}{d\mu}=\frac{-1}{\sigma}$. So we find \begin{eqnarray}
\frac{d\nu_{0}^{-1}}{d\mu} & = & \frac{\tau\sqrt{\pi}}{\sigma}\left(e^{\yr^{2}}(\erf(\yr)+1)-e^{\yth^{2}}(\erf(\yth)+1)\right)-\frac{A\tau}{\sigma}e^{-\yth^{2}}+\tau\sqrt{\pi}\frac{1}{2}(\erf(\yth)+1)\frac{dA}{d\mu}.\label{eq:dnu_dmu}\end{eqnarray}
To calculate $\frac{\partial A}{\partial\mu}$ we use \eqref{eq:flux_explicit}

\begin{align*}
\frac{1}{\tau\nu_{e}}= & \int_{\yth-\frac{w}{\sigma}}^{y_{\theta}}\Q(u)\, du.\end{align*}
Differentiating with respect to $\mu$ leads to\begin{eqnarray}
0 & = & \frac{-1}{\sigma}\left(\Q(\yth)-\Q(\yth-\frac{w}{\sigma})+\int_{\yth-\frac{w}{\sigma}}^{\yth}\frac{\partial\Q}{\partial\mu}(u)\, du\right).\label{eq:dA_dmu_first}\end{eqnarray}
The last term's integrand unfolds to\begin{eqnarray*}
\frac{\partial\Q}{\partial\mu}(u) & = & \frac{\partial A}{\partial\mu}\Qh(u)+A\underbrace{\frac{\partial\Qh}{\partial\mu}}_{=0}+\frac{\partial\Qp}{\partial\mu}(u)\\
 & = & \frac{\partial A}{\partial\mu}e^{-u^{2}}-\frac{2}{\sigma}e^{\yth^{2}-u^{2}}=\left(\frac{\partial A}{\partial\mu}-\frac{2}{\sigma}e^{\yth^{2}}\right)e^{-u^{2}},\end{eqnarray*}
where we used the explicit form \eqref{eq:Q_h} and \eqref{eq:Q_p}
of the continuous homogeneous and particular solution, respectively.
Therefore the integral in the last term of \eqref{eq:dA_dmu_first}
becomes\begin{eqnarray*}
\int_{\yth-\frac{w}{\sigma}}^{\yth}\frac{\partial\Q}{\partial\mu}(u)\, du & = & \sqrt{\pi}\left(\frac{1}{2}\frac{\partial A}{\partial\mu}-\frac{1}{\sigma}e^{\yth^{2}}\right)\left(\erf(\yth)-\erf(\yth-\frac{w}{\sigma})\right).\end{eqnarray*}
Using this expression, we can solve \eqref{eq:dA_dmu_first} for $\frac{\partial A}{\partial\mu}$
which yields\begin{eqnarray*}
\frac{\partial A}{\partial\mu} & = & \frac{2}{\sqrt{\pi}\sigma}\frac{\Q(\yth)-\Q(\yth-\frac{w}{\sigma})}{\erf(\yth)-\erf(\yth-\frac{w}{\sigma})}+\frac{2}{\sigma}e^{\yth^{2}}.\end{eqnarray*}
Inserting this result into \eqref{eq:dnu_dmu_explicit} we obtain\begin{eqnarray*}
\frac{d\nu_{0}^{-1}}{d\mu} & = & \frac{\tau\sqrt{\pi}}{\sigma}\left(e^{\yr^{2}}(\erf(\yr)+1)-e^{\yth^{2}}(\erf(\yth)+1)\right)-\frac{A\tau}{\sigma}e^{-\yth^{2}}\\
 &  & +\tau\sqrt{\pi}\frac{1}{2}(\erf(\yth)+1)\left(\frac{2}{\sqrt{\pi}\sigma}\frac{\Q(\yth)-\Q(\yth-\frac{w}{\sigma})}{\erf(\yth)-\erf(\yth-\frac{w}{\sigma})}+\frac{2}{\sigma}e^{\yth^{2}}\right)\\
 & = & \frac{\tau\sqrt{\pi}}{\sigma}e^{\yr^{2}}(\erf(\yr)+1)-\frac{\tau}{\sigma}\Q(\yth)+\frac{\tau}{\sigma}(\erf(\yth)+1)\left(\frac{\Q(\yth)-\Q(\yth-\frac{w}{\sigma})}{\erf(\yth)-\erf(\yth-\frac{w}{\sigma})}\right)\\
 & = & \frac{\tau\sqrt{\pi}}{\sigma}e^{\yr^{2}}\erfc(-\yr)-\frac{\tau}{\sigma}\Q(\yth)+\frac{\tau}{\sigma}\erfc(-\yth)\left(\frac{\Q(\yth)-\Q(\yth-\frac{w}{\sigma})}{\erf(\yth)-\erf(\yth-\frac{w}{\sigma})}\right).\end{eqnarray*}
Taken together the derivative unfolds to

\begin{eqnarray}
\frac{d\nuo}{d\mu} & = & -\nuo^{2}\frac{\tau}{\sigma}\bigg(\sqrt{\pi}e^{\yr^{2}}\erfc(-\yr)-\Q(\yth)+\erfc(-\yth)\left(\frac{\Q(\yth)-\Q(\yth-\frac{w}{\sigma})}{\erf(\yth)-\erf(\yth-\frac{w}{\sigma})}\right)\bigg).\label{eq:dnu_dmu_explicit}\end{eqnarray}

\begin{figure}
\centering{}\includegraphics{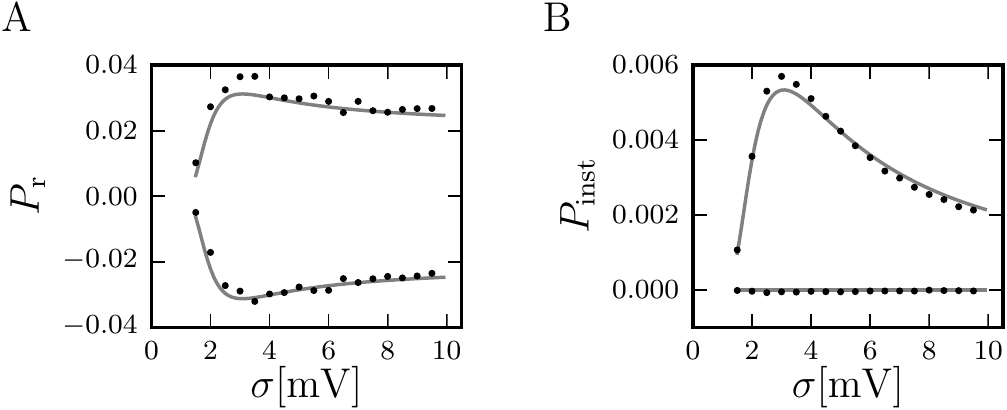}\caption{A Integral response of the firing rate \eqref{eq:int_response} of
an integrate-and-fire neuron as a function of synaptic background
noise $\sigma$. The upper trace is the response to a positive perturbation
of magnitude $s=0.5\mV$, the lower to a perturbation of $s=-0.5\mV$.
Black dots: direct simulation. Gray solid line: analytic result using
\eqref{eq:dnu_dmu_explicit}. B Instantaneous response depending on
synaptic background noise $\sigma$. The upper trace is the response
to a positive impulse of weight $s=0.5\mV$, the lower one to a negative
of $s=-0.5\mV$. Black dots: direct simulation. Gray solid line: analytic
peak response $\pinst$ using \eqref{eq:P_inst_series}. Simulations
averaged over $N=1000$ neurons for $T=100\s$. Incoming spike rates
$\nu_{e},\nu_{i}$ chosen to realize $\mu=11.5\mV$ and $\sigma$
on the abscissa for synaptic weights $w=0.1\mV$ and $g=4$. \label{fig:stochastic_resonance}}

\end{figure}

\begin{figure}
\begin{centering}
\includegraphics{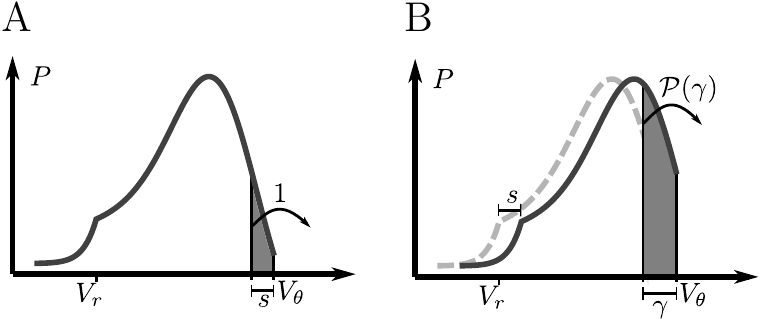}
\par\end{centering}

\caption{A Sketch illustrating the instantaneous response due to a perturbation
of size $s$: The probability mass in the shaded area is shifted above
threshold with probability $1$ and produces the instantaneous response.
Due to the shape of the probability density near the threshold, this
area has a linear and a quadratic increasing contribution in $s$.
B For all neurons in the population which have not been carried over
the threshold directly by the perturbation, the membrane potential
is shifted by $s$. During a small time interval $h\ll\tau$ after
the perturbation, the density remains shifted. This causes the firing
rate to increase approximately proportional to $-s\frac{\partial P}{\partial V}(\Vth)$
due to jumps $\gamma$ occurring with probability $\mathcal{P}(\gamma)$
by synaptic input. }
\label{fig:direct_response}
\end{figure}

To calculate the instantaneous response of the neuron to an impulse-like
perturbation we have to leave linear response theory. The incoming
spike has a direct contribution on the firing of the neuron, due to
the probability density just below threshold which is shifted over
the threshold. This is illustrated in Figure \ref{fig:direct_response}A.
This direct contribution manifests itself in a finite response probability
given by $\pinst(s)$ \eqref{eq:P_inst}, which is zero for $s<0$.
The instantaneous response has several interesting properties: For
small strength $s$ it can be expressed by the value and the first
derivative of the membrane potential distribution at the threshold.
The expression has a linear and a quadratic term in the perturbation
$s$, so the neuron will inherently respond non-linearly to a positive
perturbation $s>0$, whereas it will not respond to a negative perturbation
$s<0$, as shown in Figure \ref{fig:stochastic_resonance}B.

The integral response \eqref{eq:int_response} as well as the instantaneous
response \eqref{eq:P_inst} both exhibit stochastic resonance, an
optimal level of synaptic background noise $\sigma$ that alleviates
the transient. Figure \ref{fig:stochastic_resonance}A shows this
noise level to be at about $\sigma=3\mV$ for the integral response.
The responses to positive and negative perturbations are symmetric
and the maximum is relatively broad. The instantaneous response in
Figure \ref{fig:stochastic_resonance}B displays a pronounced peak
at a similar value for $\sigma$. This non-linear response only exists
for positive perturbations and is zero for negative ones. Stochastic
resonance has been reported for the linear response to sinusoidal
periodic stimulation \cite{Lindner01_2934} and also for non-periodic
signals that are slow compared to the neuron's dynamics an adiabatic
approximation reveals stochastic resonance \cite{Collins96c}. In
contrast to the latter work cited, the rate transient observed in
our work is the instantaneous response to a very fast (Dirac $\delta$)
perturbation.

For those neurons, which after the perturbation have not fired instantaneously,
the membrane potential is shifted by $s$. Figure \ref{fig:direct_response}B
shows a sketch of the situation just after the perturbation has been
received. The emission rate $\nu_{+}$ of the neuron in a small time
bin $(0,h]$ of size $h\ll\tau$ following the perturbation can be
approximated by similar arguments as $\pinst$. Knowing the distribution
$\Jgamma=\sum_{i,j=0}^{\infty}\mathcal{P}(i|\nu_{e}h)\mathcal{P}(j|\nu_{i}h)\delta_{w(i-gj),\gamma}$
of the membrane potential jumps $\gamma$ due to synaptic input in
a time step $h$, where $\mathcal{P}(k|\mu)=\frac{\mu^{k}}{k!}e^{-\mu}$
is the Poisson distribution, the instantaneous rate of the neuron
can be approximated by the probability per time moved above the threshold
\begin{eqnarray}
\nu_{+}(s) & = & \frac{\nuo\tau}{h}\sum_{\gamma+s>0}\Jgamma\int_{y_{\theta}-(\gamma+s)/\sigma}^{y_{\theta}-\max(s,0)/\sigma}\Q(y)\, dy\label{eq:nu_inst_null_plus}\end{eqnarray}
where the shifted density after the perturbation was inserted. Here
again we can use the expansion of $Q$ near the threshold to replace
the integral similar as in \eqref{eq:P_inst_series}\begin{eqnarray*}
\nu_{+}(s) & = & \frac{\nuo\tau}{h}\sum_{\gamma+s>0}\Jgamma\int_{y_{\theta}-(\gamma+s)/\sigma}^{y_{\theta}-\max(s,0)/\sigma}\sum_{n=0}^{\infty}\frac{\Q^{(n)}(\yth)}{n!}(y-\yth)^{n}\, dy\\
 & = & \frac{\nuo\tau}{h}\sum_{\gamma+s>0}\Jgamma\sum_{n=0}^{\infty}\frac{\Q^{(n)}(\yth)}{n+1!}\left[(-\max(s,0)/\sigma)^{n+1}-(-(\gamma+s)/\sigma)^{n+1}\right]\\
 & = & \frac{\nuo\tau}{h}\sum_{\gamma+s>0}\Jgamma\sum_{n=0}^{\infty}\frac{c_{n}+d_{n}q(\yth)}{n+1!}(-\sigma)^{-n-1}\left[\max(s,0){}^{n+1}-(\gamma+s){}^{n+1}\right].\end{eqnarray*}

As above, for small perturbations, the second summation can be truncated
at small $n$, such that the Taylor series approximates the probability
density well below threshold on the order of a few synaptic weights.
In this work we used $n=3$ throughout.